

%
%


\def\famname{
 \textfont0=\textrm \scriptfont0=\scriptrm
 \scriptscriptfont0=\sscriptrm
 \textfont1=\textmi \scriptfont1=\scriptmi
 \scriptscriptfont1=\sscriptmi
 \textfont2=\textsy \scriptfont2=\scriptsy \scriptscriptfont2=\sscriptsy
 \textfont3=\textex \scriptfont3=\textex \scriptscriptfont3=\textex
 \textfont4=\textbf \scriptfont4=\scriptbf \scriptscriptfont4=\sscriptbf
 \skewchar\textmi='177 \skewchar\scriptmi='177
 \skewchar\sscriptmi='177
 \skewchar\textsy='60 \skewchar\scriptsy='60
 \skewchar\sscriptsy='60
 \def\rm{\fam0 \textrm} \def\bf{\fam4 \textbf}}
\def\sca#1{scaled\magstep#1} \def\scah{scaled\magstephalf} 
\def\twelvepoint{
 \font\textrm=cmr12 \font\scriptrm=cmr8 \font\sscriptrm=cmr6
 \font\textmi=cmmi12 \font\scriptmi=cmmi8 \font\sscriptmi=cmmi6 
 \font\textsy=cmsy10 \sca1 \font\scriptsy=cmsy8
 \font\sscriptsy=cmsy6
 \font\textex=cmex10 \sca1
 \font\textbf=cmbx12 \font\scriptbf=cmbx8 \font\sscriptbf=cmbx6
 \font\it=cmti12
 \font\sectfont=cmbx12 \sca1
 \font\sectmath=cmmib10 \sca2
 \font\sectsymb=cmbsy10 \sca2
 \font\refrm=cmr10 \scah \font\refit=cmti10 \scah
 \font\refbf=cmbx10 \scah
 \def\twelverm{\textrm} \def\twelveit{\it} \def\twelvebf{\textbf}
 \famname \textrm 
 \advance\voffset by .06in \advance\hoffset by .28in
 \normalbaselineskip=17.5pt plus 1pt \baselineskip=\normalbaselineskip
 \parindent=21pt
 \setbox\strutbox=\hbox{\vrule height10.5pt depth4pt width0pt}}


\catcode`@=11

{\catcode`\'=\active \def'{{}^\bgroup\prim@s}}

\def\screwcount{\alloc@0\count\countdef\insc@unt}   
\def\screwdimen{\alloc@1\dimen\dimendef\insc@unt} 
\def\screwbox{\alloc@4\box\chardef\insc@unt}

\catcode`@=12


\overfullrule=0pt			
\vsize=9in \hsize=6in
\lineskip=0pt				
\abovedisplayskip=1.2em plus.3em minus.9em 
\belowdisplayskip=1.2em plus.3em minus.9em	
\abovedisplayshortskip=0em plus.3em	
\belowdisplayshortskip=.7em plus.3em minus.4em	
\parindent=21pt
\setbox\strutbox=\hbox{\vrule height10.5pt depth4pt width0pt}
\def\makefootline{\baselineskip=30pt \line{\the\footline}}
\footline={\ifnum\count0=1 \hfil \else\hss\twelverm\folio\hss \fi}
\pageno=1


\def\put(#1,#2)#3{\screwdimen\unit  \unit=1in
	\vbox to0pt{\kern-#2\unit\hbox{\kern#1\unit
	\vbox{#3}}\vss}\nointerlineskip}


\def\\{\hfil\break}
\def\newpage{\vfill\eject}
\def\center{\leftskip=0pt plus 1fill \rightskip=\leftskip \parindent=0pt
 \def\textindent##1{\par\hangindent21pt\footrm\noindent\hskip21pt
 \llap{##1\enspace}\ignorespaces}\par}
\def\unnarrower{\leftskip=0pt \rightskip=\leftskip}


\def\vol#1 {{\refbf#1} }		 


\def\NP #1 {{\refit Nucl. Phys.} {\refbf B{#1}} }
\def\PL #1 {{\refit Phys. Lett.} {\refbf{#1}} }
\def\PR #1 {{\refit Phys. Rev. Lett.} {\refbf{#1}} }
\def\PRD #1 {{\refit Phys. Rev.} {\refbf D{#1}} }


\hyphenation{pre-print}
\hyphenation{quan-ti-za-tion}

%
%


\def\oonoo#1#2#3{\vbox{\ialign{##\crcr
	\hfil\hfil\hfil{$#3{#1}$}\hfil\crcr\noalign{\kern1pt\nointerlineskip}
	$#3{#2}$\crcr}}}
\def\oon#1#2{\mathchoice{\oonoo{#1}{#2}{\displaystyle}}
	{\oonoo{#1}{#2}{\textstyle}}{\oonoo{#1}{#2}{\scriptstyle}}
	{\oonoo{#1}{#2}{\scriptscriptstyle}}}
\def\dt#1{\oon{\hbox{\bf .}}{#1}}  
\def\ddt#1{\oon{\hbox{\bf .\kern-1pt.}}#1}    
\def\slap#1#2{\setbox0=\hbox{$#1{#2}$}
	#2\kern-\wd0{\hfuzz=1pt\hbox to\wd0{\hfil$#1{/}$\hfil}}}
\def\sla#1{\mathpalette\slap{#1}}                
\def\bop#1{\setbox0=\hbox{$#1M$}\mkern1.5mu
	\lower.02\ht0\vbox{\hrule height0pt depth.06\ht0
	\hbox{\vrule width.06\ht0 height.9\ht0 \kern.9\ht0
	\vrule width.06\ht0}\hrule height.06\ht0}\mkern1.5mu}
\def\bo{{\mathpalette\bop{}}}                        
\def~{\widetilde} 
\mathcode`\*="702A                  
\def\in{\relax\ifmmode\mathchar"3232\else{\refit in\/}\fi} 
\def\f#1#2{{\textstyle{#1\over#2}}}	   
\def\half{{\textstyle{1\over{\raise.1ex\hbox{$\scriptstyle{2}$}}}}}

\def\Gamma{\mathchar"0100}
\def\Delta{\mathchar"0101}
\def\Theta{\mathchar"0102}
\def\Lambda{\mathchar"0103}
\def\Xi{\mathchar"0104}
\def\Pi{\mathchar"0105}
\def\Sigma{\mathchar"0106}
\def\Upsilon{\mathchar"0107}
\def\Phi{\mathchar"0108}
\def\Psi{\mathchar"0109}
\def\Omega{\mathchar"010A}

\catcode128=13 \def €{\"A}                 
\catcode129=13 \def {\AA}                 
\catcode130=13 \def '{\c}           	   
\catcode131=13 \def ƒ{\'E}                   
\catcode132=13 \def "{\~N}                   
\catcode133=13 \def …{\"O}                 
\catcode134=13 \def †{\"U}                  
\catcode135=13 \def ‡{\'a}                  
\catcode136=13 \def ˆ{\`a}                   
\catcode137=13 \def ‰{\^a}                 
\catcode138=13 \def Š{\"a}                 
\catcode139=13 \def ‹{\~a}                   
\catcode140=13 \def Œ{\alpha}            
\catcode141=13 \def {\chi}                
\catcode142=13 \def Ž{\'e}                   
\catcode143=13 \def {\`e}                    
\catcode144=13 \def {\^e}                  
\catcode145=13 \def '{\"e}                
\catcode146=13 \def '{\'\i}                 
\catcode147=13 \def "{\`\i}                  
\catcode148=13 \def "{\^\i}                
\catcode149=13 \def •{\"\i}                
\catcode150=13 \def –{\~n}                  
\catcode151=13 \def —{\'o}                 
\catcode152=13 \def ˜{\`o}                  
\catcode153=13 \def ™{\^o}                
\catcode154=13 \def š{\"o}                 
\catcode155=13 \def ›{\~o}                  
\catcode156=13 \def œ{\'u}                  
\catcode157=13 \def {\`u}                  
\catcode158=13 \def ž{\^u}                
\catcode159=13 \def Ÿ{\"u}                
\catcode160=13 \def  {\tau}               
\catcode161=13 \mathchardef ¡="2203     
\catcode162=13 \def ¢{\oplus}           
\catcode163=13 \def £{\relax\ifmmode\to\else\itemize\fi} 
\catcode164=13 \def ¤{\subset}	  
\catcode165=13 \def ¥{\infty}           
\catcode166=13 \def ¦{\mp}                
\catcode167=13 \def §{\sigma}           
\catcode168=13 \def ¨{\rho}               
\catcode169=13 \def ©{\gamma}         
\catcode170=13 \def ª{\leftrightarrow} 
\catcode171=13 \def «{\relax\ifmmode\acute\else\expandafter\'\fi}
\catcode172=13 \def ¬{\relax\ifmmode\expandafter\ddt\else\expandafter\"\fi}
\catcode173=13 \def ­{\equiv}            
\catcode174=13 \def ®{\approx}          
\catcode175=13 \def ¯{\Omega}          
\catcode176=13 \def °{\otimes}          
\catcode177=13 \def ±{\ne}                 
\catcode178=13 \def ²{\le}                   
\catcode179=13 \def ³{\ge}                  
\catcode180=13 \def ´{\upsilon}          
\catcode181=13 \def µ{\mu}                
\catcode182=13 \def ¶{\delta}             
\catcode183=13 \def ·{\epsilon}          
\catcode184=13 \def ¸{\Pi}                  
\catcode185=13 \def ¹{\pi}                  
\catcode186=13 \def º{\beta}               
\catcode187=13 \def »{\partial}           
\catcode188=13 \def ¼{\nobreak\ }       
\catcode189=13 \def ½{\zeta}               
\catcode190=13 \def ¾{\sim}                 
\catcode191=13 \def ¿{\omega}           
\catcode192=13 \def À{\dt}                     
\catcode193=13 \def Á{\gets}                
\catcode194=13 \def Â{\lambda}           
\catcode195=13 \def Ã{\nu}                   
\catcode196=13 \def Ä{\phi}                  
\catcode197=13 \def Å{\xi}                     
\catcode198=13 \def Æ{\psi}                  
\catcode199=13 \def Ç{\int}                    
\catcode200=13 \def È{\oint}                 
\catcode201=13 \def É{\relax\ifmmode\cdot\else\vol\fi}    
\catcode202=13 \def Ê{\relax\ifmmode\,\else\thinspace\fi}
\catcode203=13 \def Ë{\`A}                      
\catcode204=13 \def Ì{\~A}                      
\catcode205=13 \def Í{\~O}                      
\catcode206=13 \def Î{\Theta}              
\catcode207=13 \def Ï{\theta}               
\catcode208=13 \def Ð{\relax\ifmmode\bar\else\expandafter\=\fi}
\catcode209=13 \def Ñ{\overline}             
\catcode210=13 \def Ò{\langle}               
\catcode211=13 \def Ó{\relax\ifmmode\{\else\ital\fi}      
\catcode212=13 \def Ô{\rangle}               
\catcode213=13 \def Õ{\}}                        
\catcode214=13 \def Ö{\sla}                      
\catcode215=13 \def ×{\relax\ifmmode\check\else\expandafter\v\fi}
\catcode216=13 \def Ø{\"y}                     
\catcode217=13 \def Ù{\"Y}  		    
\catcode218=13 \def Ú{\Leftarrow}       
\catcode219=13 \def Û{\Leftrightarrow}       
\catcode220=13 \def Ü{\relax\ifmmode\Rightarrow\else\sect\fi}
\catcode221=13 \def Ý{\sum}                  
\catcode222=13 \def Þ{\prod}                 
\catcode223=13 \def ß{\widehat}              
\catcode224=13 \def à{\pm}                     
\catcode225=13 \def á{\nabla}                
\catcode226=13 \def â{\quad}                 
\catcode227=13 \def ã{\in}               	
\catcode228=13 \def ä{\star}      	      
\catcode229=13 \def å{\sqrt}                   
\catcode230=13 \def æ{\^E}			
\catcode231=13 \def ç{\Upsilon}              
\catcode232=13 \def è{\"E}    	   	 
\catcode233=13 \def é{\`E}               	  
\catcode234=13 \def ê{\Sigma}                
\catcode235=13 \def ë{\Delta}                 
\catcode236=13 \def ì{\Phi}                     
\catcode237=13 \def í{\`I}        		   
\catcode238=13 \def î{\iota}        	     
\catcode239=13 \def ï{\Psi}                     
\catcode240=13 \def ð{\times}                  
\catcode241=13 \def ñ{\Lambda}             
\catcode242=13 \def ò{\cdots}                
\catcode243=13 \def ó{\^U}			
\catcode244=13 \def ô{\`U}    	              
\catcode245=13 \def õ{\bo}                       
\catcode246=13 \def ö{\relax\ifmmode\hat\else\expandafter\^\fi}
\catcode247=13 \def÷{\relax\ifmmode\tilde\else\expandafter\~\fi}
\catcode248=13 \def ø{\ll}                         
\catcode249=13 \def ù{\gg}                       
\catcode250=13 \def ú{\eta}                      
\catcode251=13 \def û{\kappa}                  
\catcode252=13 \def ü{\half}     		 
\catcode253=13 \def ý{\Gamma} 		
\catcode254=13 \def þ{\Xi}   			
\catcode255=13 \def ÿ{\relax\ifmmode{}^{\dagger}{}\else\dag\fi}


\def\ital#1Õ{{\it#1\/}}	     
\def\un#1{\relax\ifmmode\underline#1\else $\underline{\hbox{#1}}$
	\relax\fi}

\def\roonoo#1#2#3{\vbox{\ialign{##\crcr
	\hfil{$#3{#1}$}\hfil\crcr\noalign{\kern1pt\nointerlineskip}
	$#3{#2}$\crcr}}}

\def\tdt#1{\oon{\hbox{\bf .\kern-1pt.\kern-1pt.}}#1}   
\def\({\eqno(}
\def\li{\openup1\jot \eqalignno}


\def\õ#1{
	\screwcount\num
	\num=1
	\screwdimen\downsy
	\downsy=-1.5ex
	\mkern-3.5mu
	õ
	\loop
	\ifnum\num<#1
	\llap{\raise\num\downsy\hbox{$õ$}}
	\advance\num by1
	\repeat}
\def\upõ#1#2{\screwcount\numup
	\numup=#1
	\advance\numup by-1
	\screwdimen\upsy
	\upsy=.75ex
	\mkern3.5mu
	\raise\numup\upsy\hbox{$#2$}}



\newcount\marknumber	\marknumber=1
\newcount\countdp \newcount\countwd \newcount\countht 

%
%
\ifx\pdfoutput\undefined
\def\rgboo#1{}
\input epsf

\def\postscript#1{\special{" #1}}		
\postscript{
	/bd {bind def} bind def
	/fsd {findfont exch scalefont def} bd
	/sms {setfont moveto show} bd
	/ms {moveto show} bd
	/pdfmark where		
	{pop} {userdict /pdfmark /cleartomark load put} ifelse
	[ /PageMode /UseOutlines		
	/DOCVIEW pdfmark}
\def\bookmark#1#2{\postscript{		
	[ /Dest /MyDest\the\marknumber /View [ /XYZ null null null ] /DEST pdfmark
	[ /Title (#2) /Count #1 /Dest /MyDest\the\marknumber /OUT pdfmark}%
	\advance\marknumber by1}
\def\pdfklink#1#2{%
	\hskip-.25em\setbox0=\hbox{#1}%
		\countdp=\dp0 \countwd=\wd0 \countht=\ht0%
		\divide\countdp by65536 \divide\countwd by65536%
			\divide\countht by65536%
		\advance\countdp by1 \advance\countwd by1%
			\advance\countht by1%
		\def\linkdp{\the\countdp} \def\linkwd{\the\countwd}%
			\def\linkht{\the\countht}%
	\postscript{
		[ /Rect [ -1.5 -\linkdp.0 0\linkwd.0 0\linkht.5 ] 
		/Border [ 0 0 0 ]
		/Action << /Subtype /URI /URI (#2) >>
		/Subtype /Link
		/ANN pdfmark}{\rgb{1 0 0}{#1}}}
%
%
\else
\def\rgboo#1{\pdfliteral{#1 rg #1 RG}}

\pdfcatalog{/PageMode /UseOutlines}		
\def\bookmark#1#2{
	\pdfdest num \marknumber xyz
	\pdfoutline goto num \marknumber count #1 {#2}
	\advance\marknumber by1}
\def\pdfklink#1#2{%
	\noindent\pdfstartlink user
		{/Subtype /Link
		/Border [ 0 0 0 ]
		/A << /S /URI /URI (#2) >>}{\rgb{1 0 0}{#1}}%
	\pdfendlink}
\fi

\def\rgbo#1#2{\rgboo{#1}#2\rgboo{0 0 0}}
\def\rgb#1#2{\mark{#1}\rgbo{#1}{#2}\mark{0 0 0}}
\def\pdflink#1{\pdfklink{#1}{#1}}
\def\xxxlink#1{\pdfklink{#1}{http://arXiv.org/abs/#1}}

\catcode`@=11

\def\wlog#1{}	


\def\makeheadline{\vbox to\z@{\vskip-36.5\p@
	\line{\vbox to8.5\p@{}\the\headline%
	\ifnum\pageno=\z@\rgboo{0 0 0}\else\rgboo{\topmark}\fi%
	}\vss}\nointerlineskip}
\headline={
	\ifnum\pageno=\z@
		\hfil
	\else
		\ifnum\pageno<\z@
			\ifodd\pageno
				\tenrm\romannumeral-\pageno\hfil\lefthead\hfil
			\else
				\tenrm\hfil\righthead\hfil\romannumeral-\pageno
			\fi
		\else
			\ifodd\pageno
				\tenrm\hfil\righthead\hfil\number\pageno
			\else
				\tenrm\number\pageno\hfil\lefthead\hfil
			\fi
		\fi
	\fi}

\catcode`@=12

\def\righthead{\hfil} \def\lefthead{\hfil}
\nopagenumbers


\def\chrulefill{\rgb{1 0 0}{\hrulefill}}
\def\cdotfill{\rgb{1 0 0}{\dotfill}}
\newcount\area	\area=1
\newcount\cross	\cross=1
\def\volume#1\par{\newpage\noindent{\biggest{\rgb{1 .5 0}{#1}}}
	\par\nobreak\bigskip\medskip\area=0}
\def\chapskip{\par\ifnum\area=0\bigskip\medskip\goodbreak
	\else\newpage\fi}
\def\chapy#1{\area=1\cross=0
	\xdef\lefthead{\rgbo{1 0 .5}{#1}}\vbox{\biggerer\offinterlineskip
	\line{\chrulefill¼\hphantom{\lefthead}\chrulefill}
	\line{\chrulefill¼\lefthead\chrulefill}}\par\nobreak\medskip}
\def\chap#1\par{\chapskip\bookmark3{#1}\chapy{#1}}
\def\sectskip{\par\ifnum\cross=0\bigskip\medskip\goodbreak
	\else\newpage\fi}
\def\secty#1{\cross=1
	\xdef\righthead{\rgbo{1 0 1}{#1}}\vbox{\bigger\offinterlineskip
	\line{\cdotfill¼\hphantom{\righthead}\cdotfill}
	\line{\cdotfill¼\righthead\cdotfill}}\par\nobreak\medskip}
\def\sect#1 #2\par{\sectskip\bookmark{#1}{#2}\secty{#2}}
\def\subsectskip{\par\ifdim\lastskip<\medskipamount
	\bigskip\medskip\goodbreak\else\nobreak\fi}
\def\subsecty#1{\noindent{\sectfont{\rgbo{.5 0 1}{#1}}}\par\nobreak\medskip}
\def\subsect#1\par{\subsectskip\bookmark0{#1}\subsecty{#1}}
\long\def\x#1 #2\par{\hangindent2\parindent%
\mark{0 0 1}\rgboo{0 0 1}{\bf Exercise #1}\\#2%
\par\rgboo{0 0 0}\mark{0 0 0}}
\def\refs{\bigskip\noindent{\bf \rgbo{0 .5 1}{REFERENCES}}\par\nobreak\medskip
	\frenchspacing \parskip=0pt \refrm \baselineskip=1.23em plus 1pt
	\def\ital##1Õ{{\refit##1\/}}}
\long\def\twocolumn#1#2{\hbox to\hsize{\vtop{\hsize=2.9in#1}
	\hfil\vtop{\hsize=2.9in #2}}}


\twelvepoint
\font\bigger=cmbx12 \sca2
\font\biggerer=cmb10 \sca5
\font\biggest=cmssdc10 scaled 3583
 \sca5

 \sca3


\def Ü{\relax\ifmmode\Rightarrow\else\expandafter\subsect\fi}
\def Û{\relax\ifmmode\Leftrightarrow\else\expandafter\sect\fi}
\def Ú{\relax\ifmmode\Leftarrow\else\expandafter\chap\fi}

\def\itemize#1 {\item{\bf#1}}
\def\itemizze#1 {\itemitem{\bf#1}}
\def\itemutem{\par\indent\indent \hangindent3\parindent \textindent}
\def\itemizzze#1 {\itemutem{\bf#1}}
\def ª{\relax\ifmmode\leftrightarrow\else\itemizze\fi}
\def Á{\relax\ifmmode\gets\else\itemizzze\fi}

\def\¢{\ominus}
\def\A{{\cal A}}  \def\B{{\cal B}}    
         \def\I{{\cal I}}
      \def\M{{\cal M}}   \def\N{{\cal N}}  
\def\O{{\cal O}}

\def\Ä{\varphi}  \def\¿{\varpi}

\def ò{\relax\ifmmode\cdots\else\dotfill\fi}


\def\cvrule{\rgbo{0 .5 1}{\vrule}}
\def\chrule{\rgbo{0 .5 1}{\hrule}}
\def\boxit#1{\leavevmode\thinspace\hbox{\cvrule\vtop{\vbox{\chrule%
	\vskip3pt\kern1pt\hbox{\vphantom{\bf/}\thinspace\thinspace%
	{\bf#1}\thinspace\thinspace}}\kern1pt\vskip3pt\chrule}\cvrule}%
	\thinspace}
\def\Boxit#1{\noindent\vbox{\chrule\hbox{\cvrule\kern3pt\vbox{
	\advance\hsize-7pt\vskip-\parskip\kern3pt\bf#1
	\hbox{\vrule height0pt depth\dp\strutbox width0pt}
	\kern3pt}\kern3pt\cvrule}\chrule}}




\def\today{\ifcase\month\or
 January\or February\or March\or April\or May\or June\or July\or
 August\or September\or October\or November\or December\fi
 \space\number\day, \number\year}

\parindent=20pt
\newskip\normalparskip	\normalparskip=.7\medskipamount
\parskip=\normalparskip	



\catcode`\|=\active \catcode`\<=\active \catcode`\>=\active 
\def|{\relax\ifmmode\delimiter"026A30C \else$\mathchar"026A$\fi}
\def<{\relax\ifmmode\mathchar"313C \else$\mathchar"313C$\fi}
\def>{\relax\ifmmode\mathchar"313E \else$\mathchar"313E$\fi}


%
%
%
%
%
%
%

\def\thetitle#1#2#3#4#5{
 \def\titlefont{\biggest} \font\footrm=cmr10 \font\footit=cmti10
  \twelverm
	{\hbox to\hsize{#4 \hfill YITP-SB-#3}}\par
	\vskip.8in minus.1in {\center\baselineskip=2.2\normalbaselineskip
 {\titlefont #1}\par}{\center\baselineskip=\normalbaselineskip
 \vskip.5in minus.2in #2
	\vskip1.4in minus1.2in {\twelvebf ABSTRACT}\par}
 \vskip.1in\par
 \narrower\par#5\par\unnarrower\vskip3.5in minus3.3in\eject}
\def\paper\par#1\par#2\par#3\par#4\par#5\par{
	\thetitle{#1}{#2}{#3}{#4}{#5}} 
\def\author#1#2{#1 \vskip.1in {\twelveit #2}\vskip.1in}
\def\YITP{C. N. Yang Institute for Theoretical Physics\\
	State University of New York, Stony Brook, NY 11794-3840}
\def\WS{W. Siegel\footnote{$*$}{
	\pdflink{mailto:siegel@insti.physics.sunysb.edu}\\
	\pdfklink{http://insti.physics.sunysb.edu/\~{}siegel/plan.html}
	{http://insti.physics.sunysb.edu/\noexpand~siegel/plan.html}}}


\pageno=0

\paper

{\rgb{0 0.8 0.5}{Superconformal spaces and implications for superstrings}}

\author{M. Hatsuda\footnote{$ÿ$}{%
	\pdflink{mailto:mhatsuda@post.kek.jp}}}
	{Theory Division, 
	High Energy Accelerator Research Organization (KEK)\\
	Tsukuba, Ibaraki, 305-0801, Japan\\
	and\\ 
	Urawa University\\
	Saitama 336-0974, Japan}
\vskip.2in
\author\WS\YITP

07-31,âKEK-TH-1180

September 28, 2007

We clarify some properties of projective superspace by using a manifestly superconformal notation.  In particular, we analyze the N=2 scalar multiplet in detail, including its action, and the propagator and its super-Schwinger parameters.  The internal symmetry is taken to be noncompact (after Wick rotation), allowing boundary conditions that preserve it off shell.  Generalization to N=4 suggests the coset superspace PSU(2,2|4)/OSp(4|4) for the AdS/CFT superstring.

\pageno=2

Û0 1. Problems with extended superspace

Several useful approaches (at least for the simplest case) have been applied to the superspaces of extended supersymmetry.  Of particular interest are subspaces of the full superspaces that use only half of the anticommuting coordinates.  One case is chiral superspaces, which use anticommuting Weyl spinors of only one chirality; however, they require the corresponding antichiral superfields for unitarity, and so save nothing when constructing actions and Feynman rules.  More important are twisted chiral superspaces, which use equal numbers of Weyl spinors of each chirality (so each half of the corresponding chirality of the full superspace), which allow reality conditions, with the help of additional ``internal" coordinates.  These approaches have a few shortcomings, as accentuated by their treatment of superconformal invariance, whose relevance has become increasingly appreciated with the advent of the AdS/CFT approach [1]:

(1) Harmonic superspace [2] isn't always manifestly superconformal; its action for the conformal $N=2$  (``improved") tensor multiplet has explicit dependence on the internal coordinates [3].  This complicates expressions for general coupling to supergravity, which is best represented as conformal supergravity, which gauges the internal symmetry, coupled to a conformal action for compensating and physical matter multiplets.  (The conformal action for compensating and physical tensor multiplets is simple in projective superspace [4].)

(2) Projective superspace [5] gives a simpler description of $N=2$ superspace and its superconformal structure with fewer internal coordinates by allowing singularities on the sphere (for the SU(2) of $N=2$):  The one-dimensional internal space is described by a complex coordinate for this sphere (without the complex conjugate).  However, the prohibition of singularities near one pole of the sphere is violated by finite SU(2) transformations, which can rotate singularities there from the opposite pole.  (Infinitesimal SU(2) invariance is not a problem.)  Also, this superspace is generally embedded in the full superspace, obscuring many of its simplifications, especially in supergraph rules.

(3) The ``flag superspace approach" of [6], as applied to $N=2$, used the same independent coordinates as the corresponding projective superspace; the supercoordinates are represented by a square matrix (torsion-free, including spacetime).   But they were applied there only on shell, where they were less simple than twistor superspaces.  They were forced on shell by the assumption, also used in harmonic superspace, of regularity everywhere in the internal space:  In harmonic superspace, the reduction of the $N=2$ internal space from two to one dimensional is itself the equation of motion.

In an earlier paper [7], we explained N=4 projective/flag superspace as arising from a type of holographic (``projective lightcone") limit [8] of AdS${}_5 ð$S${}^5$ superspace [9], and mentioned these relations between the projective and flag approaches.  Here we give more detail, with some new results, and in particular explain how the holographic interpretation naturally resolves the problem of singularities.

We first reformulate the projective superspace in a representation (``analytic") where only the independent coordinates appear, making the relation to the flag superspace approach more obvious.  This also makes superconformal symmetry (and component expansions) more transparent, especially in the action and propagator.  (The original projective approach was geared toward expansion in terms of N=1 superfields.)  

We then Wick rotate the internal symmetry to make it noncompact, with the internal space as a compact coset, identified with the boundary of the group space.  The arbitrary closed curve on the $N=2$ sphere (for SU(2)) where projective superfields live (as prescribed by contour integration in the action) is then replaced with the SU(1,1)-invariant boundary of the hyperbolic plane, justifying the manipulations used previously in projective superspace.

The propagator is the linchpin tying mechanics to field theory, as it is the solution to the free wave equation.  The one that followed from the projective superspace action can be recognized as the inverse superdeterminant, expected by symmetry from the flag superspace approach [10].  We introduce super-Schwinger parameters to make it Gaussian in the supercoordinates, allowing the usual simplifications in integration in Feynman rules.  Like the original Schwinger parameters, this supersymmetric generalization is suggested by considerations of first-quantization [7].  Since Schwinger parameters automatically arise as the fifth dimension in the projective lightcone limit of AdS$_5$, these super-Schwinger parameters are naturally interpreted as additional superspace coordinates that become nondynamical in this limit; adding these to the projective superspace coordinates for $N=4$ leads to a simple, new proposal for a coset superspace for the AdS/CFT superstring.

Û3 2. Projective superspace

Ü2.1. Coset space

Conformal symmetry in D-dimensional spacetime is described by a coset G/H, where G is the conformal group SO(D,2) and H is the Lorentz group and dilatation, H=SO(D$-$1,1)SO(1,1).  The number of D-dimensional coordinates is not the dimension of the coset G/H, but half that.  This relevant half we call the ``half-coset", denoted by G/H+.  When G is compact, such spaces are ``Hermitian symmetric spaces", and the ``half" is the restriction to just the complex coordinates and not their complex conjugates; but for spacetime symmetries G is noncompact, and usually there is no complex structure.  In the example above, the complex structure would have been implied if the SO(1,1) were SO(2).  (Our identification of this space as half of a certain coset space, rather than all of another coset with a larger isotropy group, emphasizes the similarities of this construction in the various cases.)

The 4-dimensional $N$-extended superconformal group is (P)SU(2,2|N), and the half-coset describing projective superspace is U(2,2|N)/U(1,1|$\f{N}2$)$^2$+.  (The ``P" is only for $N=4$.  The $N$ can be divided unevenly for other superspaces; e.g., all in one isotropy subgroup for chiral superspace.)  The bosonic subspace is the product of SU(2,2)/SL(2,C)GL(1)+, describing four-dimensional spacetime, and the internal space SU(N)/SU($\f{N}2$)$^2$U(1)+.  For $N=2$ the 1-dimensional internal space is SU(2)/U(1)+, representing a curve on a 2-dimensional sphere; later by Wick rotation we will consider the noncompact symmetry of the internal space SU(1,1)/U(1)+, representing the symmetry-invariant boundary of the hyperbolic plane.

The group element of U(2,2|N) is denoted as
$$ z_{\A}{}^{\M}=
\pmatrix{ z_A{}^M & z_A{}^{M'} \cr z_{A'}{}^M & z_{A'}{}^{M'} \cr}
=\pmatrix{ I & v \cr 0 & I \cr}
\pmatrix{ u^{-1} & 0 \cr 0 & u' \cr}
\pmatrix{ I& 0 \cr w & I \cr}
=\pmatrix{ u^{-1} + vu'w & vu' \cr u' w & u' \cr} $$
dividing the indices into the halves corresponding to the two U(1,1|$\f{N}2$)'s.  We have a similar expression for $z^{-1}$:
$$ \openup3\jot \li{
z^{-1} & = z_{\cal M}{}^{\cal A} =
\pmatrix{ z_M{}^A & z_M{}^{A'} \cr z_{M'}{}^A & z_{M'}{}^{A'} \cr}\cr
& =
\pmatrix{ I & 0 \cr -w & I \cr}
\pmatrix{ u & 0 \cr 0 & u'{}^{-1} \cr}
\pmatrix{ I & -v \cr 0 & I \cr}
=
\pmatrix{ u & -uv \cr -wu & u'^{-1} + wuv \cr}\cr} $$
The projective coordinates $w$ are the remainder after gauging away the rest from the left of $z_\A{}^\M$ or right of $z_\M{}^\A$.

Ü2.2. Projective space

An alternative non-coset description starts with rectangular subsets of $z$ and its inverse that are still representations of the global group:  We use only $z_{A'}{}^{\M}$ and $z_{\M}{}^A$, defined to satisfy their portion of the definition of the inverse,
$$ z_{A'}{}^{\M} z_{\M}{}^B = 0 $$
This has the solution
$$ z_{A'}{}^{\M} = u' \pmatrix{ w & I \cr}¼,â
z_{\M}{}^A = \pmatrix{ I \cr -w \cr} u $$
in agreement with the above.  Then only $u$ and $u'$ need to be gauged away.  

A set of projective coordinates are defined in terms of the first rectangle as
$$ w_{M'}{}^N ­ (z_{A'}{}^{M'})^{-1}z_{A'}{}^N = \pmatrix{
x_{Àµ}{}^à & ÐÏ_{Àµ}{}^n \cr
Ï_{m'}{}^Ã & y_{m'}{}^n \cr} $$
with indices $µ,Àµ=1,2$, and $m,m'=1,...,N/2$, and $(z_{A'}{}^{M'})^{-1}z_{A'}{}^{N'}=¶_{M'}^{N'}$, or in terms of the second as
$$ w_{M'}{}^N ­ -z_{M'}{}^A(z_N{}^A)^{-1} $$

Under the global superconformal group (P)SU(2,2|N) with elements
$$ g_\N{}^\M = \pmatrix{a&c\cr b&d\cr} $$
one of the above rectangles transforms as
$$ z_{A'}{}^\M¼£¼
z_{A'}{}^\N g_\N{}^\M =
u' \pmatrix{wa+b & wc+d\cr} $$
and thus the projective coordinate transforms as
$$ w_{M'}{}^{N}¼£¼(wc+d)^{-1}(wa+b)¼. $$

This construction is the usual definition of the term ``projective" (for real, complex, or quaternionic spaces), defining a nonlinear realization of the global group as the ratio of elements of the subgroup, which together form a representation.  (A relevant example is the projective space HP(1|$\f{N}2$), the superspace used for the supersymmetric generalization [11] of the Atiyah-Drin'feld-Hitchin-Manin construction of instantons.)  Generally one needs only one of the two rectangles (and thus no constraint), but we find it useful to consider both, for reasons we now explain.

Ü2.3. Charge conjugation

A similar result can be obtained for the transformation law of the ``inverse rectangle" $z_{\M}{}^A$ in terms of the elements of the inverse matrix $g^{-1}$.  However, it's more convenient to replace these with elements of $g$ using the unitarity condition of U(2,2|N),
$$ gÿçg = ç¼,âç^2 = 1¼,âçÿ = ç $$
where $ç$ is the U(2,2|N) metric (using bars and dots to indicate complex conjugation)
$$ ç^{À{\M}\N} = \bordermatrix{
& Ã & n & ÀÃ & n'\cr
˵ & 0 & 0 & -iC_2 & 0\cr
Àm & 0 & I & 0 & 0\cr
µ & iC_2 & 0 & 0 & 0\cr
Àm' & 0 & 0 & 0 & I\cr} $$
with, e.g.,
$$ (C_2)^{µÃ} = \pmatrix{0&i\cr -i&0\cr}¼,â(I)^{Àmn} ­ ¶_m^n $$
to obtain the same transformation law under G (but not necessarily under H) for the ``charge conjugate", defining the charge conjugation operation ``$C$" by
$$ Cz ­ z^{-1}ÿç¼,âz' = zgâÜâ(Cz)' = (Cz)g $$
where we again keep only the rectangle
$$ (z^{-1}ÿç)^{ÀA\M} = Ðz_{À{\N}}{}^{ÀA}ç^{À{\N}\M} $$
This leads directly to the definition of the charge conjugate of $w$, which will be useful for constructing Hermitian objects:  Defining $Cw$ in terms of $Cz$ by taking the same ratio as for finding $w$ in terms of $z$ (using only the rectangle parts of $z$),
$$ Cw_{M'}{}^N ­ 
\left[(zÿ^{-1}ç){}^{ÀAM'}\right]^{-1} (zÿ^{-1}ç){}^{ÀAN}
=\pmatrix{
C_2 (xÿ-Ïÿyÿ^{-1}ÐÏÿ)C_2 & -iC_2 Ïÿyÿ^{-1}\cr
iyÿ^{-1}ÐÏÿC_2 & -yÿ^{-1} \cr} . $$
(The $C_2$'s are for raising and lowering spinor indices.  Note that $C_2 x$ was Hermitian in the real representation, satisfying $C_2 xÿC_2 = x$.)

The measure for the projective coordinates transforms under the superconformal (P)SU(2,2|N) as
$$ dw¼£¼dw¼[sdet (wc+d)] ^{-2¼str(I)}¼,âstr(I)=2-\f{N}2¼. $$
This follows from
$$ dz_{A'}{}^{\M}¼£¼dz_{A'}{}^{\M} $$
$$ dz = dw¼du'¼[sdet(u')]^{str(I)} $$
$$ u'¼£¼u'(wc+d)âÜâdu'¼£¼du'¼[sdet(wc+d)]^{str(I)} $$
where the power of $str(I)$ for the transformation of $du'$ comes from performing it separately for each value of $A'$ on $du'$ (and similarly for $dz$ in terms of $du'$ and $dw$).  The measure for the charge conjugated coordinates $Cw$ is related to the above through the Jacobian:
$$ [d(Cw)]ÿ = dw¼sdet\left[{»(Cw)ÿ\over »w}\right] = dw¼[det(y)]^{2¼str(I)}¼. $$
Then one defines the superfield $Æ(w)$ to be a density in terms of the scalar ``volume element"
$$ dw[Æ(w)]^2 = dw'[Æ'(w')]^2 $$
So under charge conjugation we have
$$ dw[(CÆ)(w)]^2 ­ Ód(Cw)[Æ(Cw)]^2ÕÿâÜâ(CÆ)(w) = [det(y)]^{str(I)} [Æ(Cw)]ÿ $$
Then the action
$$ S = Çdw¼(CÆ)(w) Æ(w) $$
is invariant under both superconformal transformations and charge conjugation.  For the $N=2$ scalar multiplet of the next section, the factor $[det(y)]^{str(I)}$ is simply $y$.  (This factor of $y$ was introduced as a measure factor in the original projective superspace approach.)

The existence of this type of ``complex conjugation", and the corresponding reality conditions, allows the definition of real quantities, even though one is working on a space whose coordinates are not considered real.  In fact, even with the usual real coordinates, one is forced to consider complexification even for spacetime itself, e.g., for integration around poles in Feynman diagrams in Minkowski space; in that case, one never treats this as doubling of coordinates --- the complex-conjugate coordinates are not considered.  This should be distinguished from the case of ``holomorphic fields" on the Euclidean worldsheet in string theory, since there antiholomorphic fields also exist.  This is especially clear from the fact that both string coordinates are always required to write actions.  In contrast, here even the (real) actions are written on the half-coset.

Û3 3. Scalar hypermultiplet

Ü3.1. Transformation to analytic representation

For $N=1$ chiral superfields, the chirality condition $ÐdÄ=0$ becomes simply independence from $ÐÏ$, $лÄ=0$, by the complex coordinate transformation
$$ x¼£¼x +iÏÐÏâÜâÐd¼£¼Ð» $$
from the real representation to the chiral one.  The original formulation of projective $N=2$ superspace was in a real representation.  The constraints there also can be converted into partial derivatives by transforming to an ``analytic" representation.  (A similar transformation was applied in harmonic superspace [2].)  The choice of analytic $Ï$'s is obvious if the superspace is written in 6-dimensional notation, since all 4D $N=2$ matter multiplets exist as multiplets of minimal 6D supersymmetry.  This should not be confused with the choice, usually made by the proponents of projective superspace, of $N=1$ $Ï$'s:  Projective superspace is a twisted superspace, using a $Ï$ and $ÐÏ$ that are not complex conjugates.  In the following we will use 6D notation as shorthand:  $÷Ï$ is the 6D charge conjugate of the 4-component $Ï$; applying it twice gives $-Ï$ because of pseudoreality.  (I.e., $÷Ï$ is the complex conjugate of $Ï$ times an antisymmetric matrix.)  We also write as a 4-vector (part of a 6-vector) expressions quadratic in these spinors, where, e.g., $Ï÷Ï=+÷ÏÏ$ because this vector is antisymmetric in the spinor indices.

For $N=2$, the analyticity condition $÷d=0$ (where $÷d$ is a 4-spinor) becomes independence from $÷Ï$ (the fermionic coordinates ÓnotÕ included in $w$) by successive complex coordinate transformations:  For $$ an ``arctic" superfield in the real representation ($÷d=0$) corresponding to $Æ$ in the analytic one ($÷»Æ=0$), 
$$ ÷d = ÎX ÷» X^{-1}Î^{-1}¼,⍠= ÎX Æ $$
$$ Xx = x +Ï÷ϼ,¼ÎÏ = Ï +y÷Ï $$
from the analytic representation to the real one (where $ÎXÆ(w)=Æ(ÎXw)$, etc.; our coordinate transformation operators are exponentials of derivatives).

This transformation of coordinate representations is straightforward on any superfield, but nonunitary.  This is similar to the case for $N=1$ chiral superfields, where the transformation to the representation that removes $ÐÏ$ from chiral superfields does not perform a similar service for antichiral superfields.  The situation is not as bad in the projective case:  In the real representation usually used for projective superspace, complex conjugation must be accompanied by the replacement $yÿ£-1/y$ for the internal coordinate.  As we have seen, this is just the internal part of charge conjugation.  That is all that is required in the real representation, since there only $y$ is treated as a complex coordinate, while $x$ is real in the usual sense, and all $Ï$'s still appear.  Combining this with the transformation to the analytic representation produces the charge conjugation derived in the previous section:  The resulting action is
$$  S = Ç(\I )ÿ = Ç[Æ(Cw)]ÿÆ(w) $$
in terms of the $y$-inversion
$$ \I y = -1/y* $$ 
where $$ and $Æ$ are defined to be Taylor expandable in $y$ as boundary conditions.  The relation between the two forms of the action follows from
$$ X^{-1}Î^{-1}[(\I )ÿ] = (X*^{-1}Î*^{-1}\I ÎX Æ)ÿÆ ­ (\O Æ)ÿƼ,â\O w = Cw $$
where applying the successive transformations, we found
$$ \O y = -1/y*¼,â\O Ï = -÷Ï/y*¼,â\O x = x - ÷Ï÷Ï/y* $$
which agrees with the previous expression for $Cw$ after making the appropriate definition of the 4-spinor $Ï$ in terms of the 2-spinors $Ï$ and $ÐÏ$.  (The Jacobian determinant of $X$ and $Î$ is 1.)

The relation of the propagator in that formalism to the $1/sdet(w-w')$ required by superconformal invariance is slightly more obscure:  They wrote it as (in our notation, but still in the real representation)
$$ Ò(\I )ÿ(w) (w')Ô = y^{-1}(y-y')^{-3}÷d^4 ÷d'^4 õ^{-1}¶^4(÷Ï-÷Ï')¶^4(Ï-Ï')¶^4(x-x') $$
The $y^{-1}$ is canceled by using $CÆ$ in place of $[Æ(Cw)]ÿ$.  The transformation from the real to the analytic representation given above replaces 
$$ \li{ & ÷d^4 ÷d'^4 ¶^4(÷Ï-÷Ï')¶^4(Ï-Ï')¶^4(x-x') \cr
	£¼& ÷»^4 ÷»'^4 ¶^4(÷Ï-÷Ï')¶^4[(Ï-Ï')-(y-y')÷Ï]¶^4[(x-x')-(Ï-Ï')÷Ï] \cr
	=¼& (y-y')^4 ¶^4[(x-x')-(Ï-Ï')(ÐÏ-ÐÏ')/(y-y')] \cr} $$
where we used the first $¶$ to replace $÷Ï'£÷Ï$ in the others, then killed it with the $÷»'^4$ (which is the same as $÷Ï'$ integration), and finally used the second $¶$ with the $÷»^4$ to replace $÷Ï$.  (Then we reverted to 2-spinor notation.)  Using $õ^{-1}¶^4(x)=1/x^2$ (up to normalization), the final result is
$$ Ò(CÆ)(w) Æ(w')Ô = {y-y'\over [(x-x') -(Ï-Ï')(ÐÏ-ÐÏ')/(y-y')]^2} = {1\over sdet(w-w')} $$
(Similar results have been obtained in the harmonic superspace formalism [12], although they were not identified in terms of the superdeterminant.  Differences are expected from the fact that multiplets there live on a larger internal space off shell.)

Ü3.2. Wick rotation

In general, quantum theories are treated off shell in Euclidean space (then Wick rotated back), on shell directly in Minkowski space:  The (St¬uckelberg-Feynman) propagator for a real field is real in Euclidean space, complex in Minkowski space; so quantum fields keep their reality properties only in Euclidean space.  (Similar remarks apply to instanton solutions.)  But the wave equation for a massive field has no solutions in Euclidean space ($p^2+m^2$ is strictly positive there), so asymptotic states are defined only in Minkowski space.

However, Wick rotation of spinors, and thus fermionic coordinates, leads to problems with reality, since Majorana properties depend not only on the number of dimensions of spacetime but also on its signature.  Here we extend the question to the internal coordinates $y$ of projective superspace.  We find that Wick rotation solves a problem with finite internal symmetry transformations off shell, in a way suggested by the derivation of projective superspace from a projective lightcone (``holographic") limit.

The problem is that in the projective superspace approach superfields must have singularities in the internal space off shell.  For example, the arctic superfield has singularities in $y$ only at infinity, so it can be Taylor expanded about the origin.  This is not a problem on shell:  There the only nonvanishing components are at zeroth and first orders in $y$.  The $y$ component may seem singular at infinity, but this is only a coordinate singularity:  Using the transformation for $Æ$ given in the previous section, performing the rotation $y£-1/y$ of each pole to the other, the density transformation of $Æ$ produces an extra factor of $y$ that returns the regular form of $Æ$.  (Similar remarks apply to the $Ï$ dependence.)  But the regularity of the off-shell components is lost:  All negative powers of $y$ are produced.  (Infinitesimal SU(2) transformations will never show the motion of the singularities from infinity.)

The solution we propose is to replace the internal symmetry group SU(2) with the noncompact SU(1,1) by Wick rotation off shell:  The complex coordinate of the sphere, used for contour integration in the action, is then replaced with the coordinate of the boundary of the hyperbolic plane (2D space of constant negative curvature), which is invariant under the group.

Ü3.3. Half-cosets

In general, the metric of a space of constant (scalar) curvature in D dimensions can be written in conformally flat coordinates as
$$ {dx^2\over (a + bÉx + cüx^2)^2} $$
with curvature tensor
$$ R_{ab}{}^{cd} = -(b^2 - 2ac)¶_{[a}^c ¶_{b]}^d $$
The constants $(a,c,b_m)$ form a (D+2)-vector, in lightcone basis, that breaks conformal invariance; conformal transformations SO(D+1,1) (or a Wick rotation, as appropriate to adding 1 space and 1 time dimension to the D-dimensional space) rotate this vector, but don't change its magnitude, the curvature.  Here we will consider spaces of Euclidean signature in D=2 (used in the inner products indicated above), as for the sphere.  However, we Wick rotate to change the sign of the curvature from positive to negative, making it noncompact.

There are 2 convenient choices for the 4-vector $(a,c,b_m)$.  The choice naturally identifying the coset as SL(2,R)/SO(2) is to use a component of $b$:  Labeling the 2 coordinates $y$ and $y_0$, we find the metric
$$ {dy^2+(dy_0){}^2\over (y_0){}^2} $$
of the PoincarŽ (upper) half-plane.  Inserting back a scale into the metric gives
$$ ds^2={dy^2+R^2(dy_0){}^2\over (y_0){}^2} $$
The projective lightcone limit $R¼£¼ 0$ yields the boundary $y_0= 0$,
the real axis of the complex plane, a
one dimensional flat space with conformal factor
$$ ds^2 ={dy^2\over (y_0){}^2} $$
The $y$ coordinate is the coset parameter of the half-coset SL(2)/SO(2)+.  In these coordinates we see the close analogy to the holographic treatment of AdS$_5$.  We should also recognize the invariance of the real axis under SL(2) from its application to open-string tree amplitudes.

The choice naturally associated with SU(1,1)/U(1) is to use $a$ and $c$:  In terms of a complex coordinate $y$ and its complex conjugate $Ðy$, the resulting metric is
$$ {dy¼dÐy\over (1-yÐy)^2} $$
of the PoincarŽ disk, which is identical to the metric for the sphere except for the change in sign in the denominator.  The invariant boundary is now $|y|=1$:  We can thus choose this boundary as the contour of integration in the action, for functions nonsingular in $y$ on the disk.  This $y$ parametrizes the half-coset SU(1,1)/U(1)+.

Û3 4. Propagator

Ü4.1.  Field equations

The equations of motion for a superspinless projective multiplet are [7]
$$ »_{[M}{}^{[M'}»_{N)}{}^{N')} = 0 $$
These include the massless Klein-Gordon equation in spinor notation and the equation of $û$ symmetry, common to all massless multiplets.  They also include the analog of the Klein-Gordon equation for the internal coordinates:
$$ »_{(m}{}^{(m'}»_{n)}{}^{n')} = 0 $$
which differs in that the indices are symmetrized instead of antisymmetrized.  For $N=2$ this is simply $»^2/»y^2=0$.  Written in terms of vector indices, this equation for $N=4$ becomes
$$ »_i »_j - \f1D ú_{ij}ú^{kl}»_k »_l = 0 $$
for indices $i,j,...$ that range over $D=4$ values.  This equation is $D$-dimensionally conformal:  Just as the Klein-Gordon equation can be made Weyl scale invariant in curved space by adding a curvature term
$$ õ - \f14 \f{D-2}{D-1}R = 0 $$
so can this one as
$$ á_i á_j + \f1{D-2}R_{ij} - tr = 0 $$

Since the linearly realized superconformal transformations include translations and (P)SU(1,1|$\f{N}2$)$^2$ transformations, the only candidate for a Green function is some function of the superdeterminant.  It's easily checked, using
$$ d[ln¼sdet(M)] = str(M^{-1}dM),âd(M^{-1}) = - M^{-1}(dM)M^{-1} $$
$$ Üâ»_M{}^{M'}[ln¼sdet(w)] = w^{-1}{}_M{}^{M'},â
	»_M{}^{M'}w^{-1}{}_N{}^{N'} = - w^{-1}{}_M{}^{N'}w^{-1}{}_N{}^{M'} $$
(where extra signs from reordering of graded indices are understood), that
$$ »_{[M}{}^{[M'}»_{N)}{}^{N')}[sdet(w-w')]^{-1} = 0âfor¼w ± w' $$
For $w=w'$, there are $¶$-function terms, as elucidated by an (Euclidean) $·$ prescription:
$$ \openup2\jot \li{ »_{[M}{}^{[M'}»_{N)}{}^{N')}{1\over sdet(w) + ·}
	& = -2·¼{sdet(w)\over [sdet(w) + ·]^3}w^{-1}{}_{[M}{}^{[M'}w^{-1}{}_{N)}{}^{N')}\cr
	& = - ¶[sdet(w)]w^{-1}{}_{[M}{}^{[M'}w^{-1}{}_{N)}{}^{N')}\cr} $$
(for $sdet(w)\ge 0$).  For example, for $N=0$ this gives the usual, since $¶(x^2)/x^2=¹^2 ¶^4(x)$ (in Euclidean space).

The propagtor is thus
$$ ë(w,w') = {1\over sdet(w-w')} $$ 
where the superdeterminant takes the usual form; for $N=2$,
$$ {1\over sdet(w)} = {y\over (x - ÏÐÏ/y)^2} $$
The fact that the $y$ determinant appears in the numerator and the $x$ determinant appears in the denominator of such propagators questions their relationship to AdS$_5ð$S$^5$, where $x$ and $y$ are treated symmetrically (at least locally).

This should be compared with the corresponding $N=1$ case, the propagator connecting a chiral scalar multiplet superfield to an antichiral one,
$$ ë(x,Ï;x',ÐÏ') = {1\over (x-x' +iÏÐÏ')^2} $$
Note that a Taylor expansion in $Ï$ of either one will first produce $1/x^2$ terms for scalars, then $1/x^3$ terms for spinors, and finally $¶^4(x)$ terms for auxiliary fields (as follows from the wave equation satisfied by the propagator).

Ü4.2. Super-Schwinger parameters

Schwinger parametrization is convenient for evaluation of Feynman diagrams as well as relating them to first-quantization.  For both these purposes, the parametrization must correspond to a positive-definite action (after Wick rotation):  Thus, the coordinates (or momenta) must appear quadratically, while the Schwinger parameter is constrained to be positive, which can be accomplished by having it also appear quadratically.

In [7] we introduced a ``quadratic multplier" action for the superparticle in projective superspace, inspired by that following from the projective lightcone limit of the AdS$_5ð$S$^5$ Green-Schwarz action, but avoiding second-class constraints, and with both the coordinates and multipliers appearing quadratically.  Ignoring questions of quantization by directly replacing the path integral with an ordinary integral, and derivatives with finite differences, the result is a Schwinger parametrization appropriate for producing the superdeterminant:
$$ Çdu¼du'¼µ¼e^{-S_2} = {1\over sdet(w-w')} $$
$$ S_2 = üú^{A'B'}ú_{AB}J_{A'}{}^A J_{B'}{}^B $$
$$ J_{A'}{}^A = u_{A'}{}^{M'}(w-w')_{M'}{}^M u_M{}^A $$
$$ µ = [sdet(u_M{}^A)sdet(u_{A'}{}^{M'})]^{N/2-1} $$
where $ú$ and $ú'$ are the metrics of OSp($\f{N}2$|2).  $u$ and $u'$ then belong to the coset (P)S[U(1,1|$\f{N}2$)$^2$]/OSp($\f{N}2$|2)$^2$.  (The ``S", and ``P" for $N=4$, deal with the fact that $u_{A'}{}^{M'}$ and $u_M{}^A$ appear only as their product.)  This action can be recognized as the quadratic multiplier action if we identify ``metrics" $e,e'$ for the ``vielbeins" $u,u'$:
$$ e^{M'N'} = ú^{A'B'}u_{A'}{}^{M'}u_{B'}{}^{N'},â
	e_{MN} = ú_{AB}u_M{}^A u_N{}^B $$
$$ ÜâS_2 = üe^{M'N'}e_{MN}(w-w')_{M'}{}^M (w-w')_{N'}{}^N $$

The integral can be evaluated by using a first-order formalism,
$$ S_1 = üú^{AB}ú_{A'B'}P_A{}^{A'}P_B{}^{B'} + iP_A{}^{A'}J_{A'}{}^A $$
This allows the $u,u'$ integrals to be evaluated without $µ$:
$$ \li{ {1\over sdet(w)}
& = Çdu¼du'¼dP¼[sdet(u)sdet(u')]^{N/2-1} e^{-P^2/2 -iPu'wu} \cr
& = [sdet(w)]^{1-N/2} ÇdP¼ e^{-P^2/2}
	[sdet( i»_P)]^{N/2-1} Çdu¼du'¼e^{-iPu'wu} \cr
& = [sdet(w)]^{-1} ÇdP¼ e^{-P^2/2}
	[sdet( i»_P)]^{N/2-1} [sdet(iP)]^{N/2-2} \cr} $$
where we have used the identity
$$ sdet(e^M) = e^{str(M)}âÜâsdet(M°N) = [sdet(M)]^{str(I_N)}[sdet(N)]^{str(I_M)} $$
as applied to $P°w$.  The final $P$ integral can then be normalized to 1.  (The answer can also be obtained using superconformal invariance to show that the result is $sdet(w)$ to a power determined by dimensional analysis, where the normalization can be fixed by considering the case $w=I$.)

The measure $µ$ is a bit arbitrary, since it depends on the choice of variables of integration:  For example, we could make changes of variables such as
$$ du¼du' = d(u^{-1})¼d(u'^{-1}) [(sdet(u)sdet(u')]^2 $$
since $u_A{}^M$ and $u_{M'}{}^{A'}$ seem as good a choice as $u_M{}^A$ and $u_{A'}{}^{M'}$.  Another choice depends on the form of the first-order formalism:
$$ P'_M{}^{N'} = u_M{}^A P_A{}^{B'}u_{B'}{}^{N'}âÜâdP' = dP [(sdet(u)sdet(u')] $$
With our choices of variables $µ=1$ for $N=2$, but with the latter change it would be trivial for $N=4$.  Alternatively, such factors might be generated by ``ghosts":  For example, for $N=4$, we could use the extra action
$$ S_µ = \f12 ½^M ½^N e_{MN} + \f12 ½_{M'}½_{N'}e^{M'N'}¼,âµ = Çd½¼d½'¼e^{-S_µ} $$
where the $½$'s have normal statistics (e.g., $½^Œ$ is an anticommuting spinor; for $N=0$ they would have opposite statistics; other powers of the $sdet(u)$'s can be generated by multiple copies of $½$'s).  Note that positivity of these terms is guaranteed by that of $e$ and $e'$, which are themselves expressed as squares of $u$ and $u'$.  (The same would not be true for using an analogous term of the form $ÂÀwÂ'£Â(w-w')Â'$ to generate the original superdeterminant, since $w-w'$ isn't positive definite:  Twistor actions of the form $ÂÀwÂ'$ imply conjugate momentum $p=ÂÂ'$ positive in Minkowski space, and hence do not lead to St¬uckelberg-Feynman propagators.)

Ü4.3.  New coset

Having rewritten the action of [7] in terms of the currents $J$ (of elsewhere in that same paper), we can now see that, although the dynamic variables $w$ belong to the same coset U(2,2|N)/U(1,1|$\f{N}2$)$^2$+ as before, the variables $w,u,u'$ collectively belong to (P)SU(2,2|N)/OSp($\f{N}2$|2)$^2$+, which differs from that obtained from the projective lightcone limit of the Green-Schwarz action.  This new coset can be obtained from the same limit of the coset (P)SU(2,2|N)/OSp(N|4), which is a more natural coset to describe the superspace of AdS$_5$, as OSp(N|4) is the supersymmetry group of AdS$_4$, so this is the direct supersymmetrization of AdS$_5$ = SO(4,2)/SO(4,1).  (Actually, it generalizes SO(4,2)/SO(3,2), so there is some question about the change in signature, and sign of the curvature.)  Also, some variables were lost in the limit for the previous coset, while for this new coset all the same variables $w,u,u'$ remain, only $u$ and $u'$ become nondynamical (in direct analogy to the $N=0$ case).

The action for this coset takes a very simple form in a block triangular gauge for OSp(N|4) (as is allowed by the graded generalization of the decomposition of a general matrix into orthogonal and triangular factors):  Gauging $v=0$ in the expression for a general element of (P)SU(2,2|N), and inserting factors of $R$ as defined by the projective lightcone limit, the action is
$$ S\left[\textstyle{(P)SU(2,2|N)\over OSp(N|4)}\right] = S_{PLC}[w,u,u'] 
	+ R^2 S\left[\textstyle{(P)S[U(1,1|N/2)^2]\over OSp(N/2|2)^2} \right] $$
where $S_{PLC}$ is the action of the previous subsection, and $R^2 S[...]$ is the action for $u$ and $u'$ (no $w$ appears there), since after gauging $v=0$, OSp($\f{N}2$|2)$^2$ is the residual gauge invariance.  Modified projective lightcone limits yield other superspaces, such as chiral, by just changing the range of the ``$M$" and ``$M'$" indices to (n|2) and (N$-$n|2) for n$±$N/2 [7].  However, these superspaces have no charge conjugation (because of unequal numbers of $Ï$'s and $ÐÏ$'s), and so are not useful for constructing field theory actions.

Because of this residual invariance, all these actions can easily be rewritten in forms where $u$ and $u'$ are replaced by $e$ and $e'$.  The same is true for the full coset (P)SU(2,2|N)/OSp(N|4):  We can replace the (P)SU(2,2|N) group element $z$ with
$$ E^{\M\N} = ú^{\A\B}z_\A{}^\M z_\B{}^\N = 
	\pmatrix{ e^{MN} + e^{P'Q'}w_{P'}{}^M w_{Q'}{}^N & e^{N'P'}w_{P'}{}^M \cr
			e^{M'P'}w_{P'}{}^N & e^{M'N'} \cr } $$
where we have substituted the triangular gauge ($v=0$) form for $z$.  $E$ satisfies the condition
$$ 1 = sdet(E^{\M\N}) = { sdet(e^{M'N'})\over sdet(e_{MN}) } 
	= \left[{ sdet(u_{A'}{}^{M'})\over sdet(u_M{}^A) }\right]^2 $$
where $e^{MN}$ is the inverse of the $e_{MN}$ that appears in $S_{PLC}$.  The action can be written as
$$ S ¾ Ç str( ÀE E^{-1} ÀE E^{-1} ) $$
which has the ``P" invariance $E£E e^{iÄ}$ as a consequence of the $sdet$ condition for the case $N=4$.  ($E$ also satisfies a unitarity condition, but this is obscured in the analytic representation, where coordinates appear without their complex conjugates.)

Û2 5. Prospects

We have reviewed projective superspace, and in particular the $N=2$ scalar multiplet, in a notation where the irrelevant coordinates of the rest of the full superspace do not appear, so that superconformal transformations are manifest.  This allowed a simple derivation of charge conjugation, and thus the action, in a new, manifestly superconformal form.  We then proposed a holographic interpretation of the integration contour that avoided singularities after finite SU(2) transformations.  Finally, we gave a manifestly superconformal Schwinger parametrization of the propagator, and found that it could be obtained as a projective lightcone limit from a new coset on an AdS$_5$ spacetime.

Our general discussion may allow generalization to the $N=4$ superparticle.  This also would require an understanding of the ghost structure, to understand the Yang-Mills gauge field and not just its field strength.  (Such behavior is known from quantization of the extended spinning particle [13].)  For example, although a superdeterminant can again be used to describe the propagator of the (super)field strength, the same is not true for gauge-dependent quantities such as the gauge field, since gauge fixing destroys manifest conformal invariance.

Our general discussion of actions in projective superspace should help generalization to $N=4$ also.  There we expect complications simply from the fact that the internal space increases dimensions from 1 to 4, allowing choices for gauge-covariant wave equations in that space from index structure alone.

The new coset suggested by super-Schwinger parametrization would require a new string action (from the case $N=4$).  The internal space is not identifiable (classically) with S$^5$, so this model would be a dual to the usual one, if the conformal anomaly cancels after including ghosts.  (Extra ``dimensions" don't hurt if they are compact:  This is known, e.g.,  from Wess-Zumino models, where currents with a given level number, and thus conformal anomaly, are represented by a boson for each generator of the algebra, rather than one for each generator of the Cartan subalgebra [14].)  The ghost structure is not obvious, since the quadratic multipliers cannot be gauge fields like the usual linear Lagrange multipliers:  They may themselves be the result of a gauge fixing for $û$ symmetry and its relatives; however, we saw that the superdeterminant propagator still satisfies the wave equations for these symmetries.  The multipliers must also appear as some type of moduli or conserved quantities (at least in the superparticle case) to allow the path integral for the propagator to reduce to an ordinary integral as above.

ÜAdded note

After this paper appeared on the archive, Kuzenko [15] described (non-manifest) superconformal transformations on projective superspace in the real representation.

ÜAcknowledgments

W.S. was supported in part by the National Science Foundation Grant No. PHY-0653342.
M.H. acknowledges the Simons Workshop in Mathematics and Physics 2007
and the CNYITP at Stony Brook for their hospitality, where part of this work was done.
M.H. was supported by the Grant-in-Aid for Scientific Research No.¼18540287.

\refs

£1 L. Maldacena, ÓAdv. Theor. Math. Phys.Õ É2 (1998) 231, \xxxlink{hep-th/9711200};\\
S.S. Gubser, I.R. Klebanov  and A.M. Polyakov, \PL 248B (1998) 105, \xxxlink{hep-th/9802109};\\
E. Witten, ÓAdv.  Theor. Math. Phys.Õ É2 (1998) 253, \xxxlink{hep-th/9802150}.

£2 A. Galperin, E. Ivanov, S. Kalitzin, V. Ogievetsky, and E. Sokatchev, 
ÓClass. Quantum Grav.Õ É1 (1984) 469;\\
A. Galperin, E. Ivanov, V. Ogievetsky, and E. Sokatchev, 
ÓJETP Lett.Õ É40 (1984) 912 (ÓPis'ma Eksp. Theor. Fiz.Õ É40 (1984) 155).

£3 A. Galperin, E. Ivanov, and V. Ogievetsky, 
ÓSov. J. Nucl. Phys.Õ É45 (1987) 157 (ÓYad. Fiz.Õ É45 (1987) 245).

£4 N. Berkovits and W. Siegel, \NP 462 (1996) 213, \xxxlink{hep-th/9510106};\\
W. Siegel, \PRD 53 (1996) 3324, \xxxlink{hep-th/9510150}.

£5 A. Karlhede, U. Lindstr¬om, and M. Ro×cek, \PL 147B (1984) 74;\\
U. Lindstr¬om and M. Ro×cek, ÓComm. Math. Phys.Õ É115 (1988) 21; 
ÓComm. Math. Phys.Õ É128 (1990) 191;\\
I.T. Ivanov and  M. Ro×cek, ÓComm. Math. Phys.Õ É182 (1996) 291;\\
F. Gonzalez-Rey, M. Ro×cek, S. Wiles, U. Lindstr¬om, and R. von Unge, 
\NP 516 (1998) 426, \xxxlink{hep-th/9710250};\\
F. Gonzalez-Rey and R. von Unge, \NP 516 (1998) 449, \xxxlink{hep-th/9711135};\\
F. Gonzalez-Rey, \xxxlink{hep-th/9712128};\\
F. Gonzalez-Rey and  M. Ro×cek, \PL 434B (1998) 303, \xxxlink{hep-th/9804010}.

£6 G.G. Hartwell and P.S. Howe, ÓInt. J. Mod. Phys.Õ É10 (1995) 3901,
\xxxlink{hep-th/9412147};\\
S. Howe and G.G. Hartwell, ÓClass. Quantum Grav.Õ É12 (1995) 1823;\\
P.J. Heslop and P.S. Howe, ÓClass. Quantum Grav.Õ É17 (2000) 3743, 
\xxxlink{hep-th/0005135};\\
P.J. Heslop, ÓClass. Quantum Grav.Õ É19 (2002) 303, \xxxlink{hep-th/0108235}.

£7 M. Hatsuda and W. Siegel, \PRD 67 (2003) 066005, \xxxlink{hep-th/0211184}.

£8 H. Nastase and W. Siegel, ÓJHEPÕ É0010 (2000) 040, \xxxlink{hep-th/0010106}.

£9 R.R. Metsaev and A.A. Tseytlin, \NP 533 (1998) 109, \xxxlink{hep-th/9805028}.

£10   P.S. Howe and P.C. West, \PL 389B (1996) 273, \xxxlink{hep-th/9607060};\\
\xxxlink{hep-th/9611074}.

£11 W. Siegel, \PRD 52 (1995) 1042, \xxxlink{hep-th/9412011}.

£12 A. Galperin, E.A. Ivanov, V. Ogievetsky, and E. Sokatchev,
ÓClass. Quant. Grav.Õ É2 (1985) 601;\\
A.S. Galperin, E.A. Ivanov, V.I. Ogievetsky, and E.S. Sokatchev, ÓHarmonic SuperspaceÕ 
(Cambridge Univ., 2001), eqs.\ (8.17-18).

£13 W. Siegel, ÓInt. J. Mod. Phys. AÕ É6 (1991) 3997.

£14 S.P. Novikov, ÓSov. Math. Dokl.Õ É24 (1981) 222, ÓUsp.\ Mat.\ NaukÕ É37N5 (1982) 3;\\
E. Witten, ÓCommun. Math. Phys.Õ É92 (1984) 455.

£15 S.M. Kuzenko, \xxxlink{arXiv:0710.1479}.

\bye